\documentclass[aps,prl,twocolumn,showpacs,10pt,superscriptaddress,preprintnumbers,nofootinbib]{revtex4-1}
\usepackage{graphicx}
\usepackage{amsmath,bm,braket}
\usepackage{color}

\allowdisplaybreaks

\newcommand{\dd}{{\mathrm d}}
\newcommand{\als}{\alpha_s}

\newcommand{\nn}{\nonumber}

\begin{document}

\preprint{SLAC-PUB-15260, SMU-HEP-12-19}

\title{Top-Quark Decay at Next-to-Next-to-Leading Order in QCD}
\author{Jun Gao}
\email{jung@smu.edu}
\affiliation{Department of Physics, Southern Methodist University, Dallas,
TX 75275-0175, USA}
\author{Chong Sheng Li}
\email{csli@pku.edu.cn}
\affiliation{Department of Physics and State Key Laboratory of Nuclear
  Physics and Technology, Peking University, Beijing 100871, China}
\affiliation{Center for High Energy Physics, Peking University,
  Beijing 100871, China}
\author{Hua Xing Zhu}
\email{hxzhu@slac.stanford.edu}
\affiliation{SLAC National Accelerator Laboratory, Stanford University,
Stanford, CA 94309, USA}

\begin{abstract}
We present the complete calculation of the top-quark decay width at next-to-next-to-leading order in QCD,
including next-to-leading electroweak corrections as well as finite bottom quark mass and $W$ boson width effects.
In particular, we also show the first results of the fully differential decay rates for
top-quark semileptonic decay $t\to W^+(l^+\nu)b$ at next-to-next-to-leading order in QCD.
Our method is based on the understanding of the invariant mass distribution of the final-state
jet in the singular limit from effective field theory.
Our result can be used to study arbitrary infrared-safe observables
of top-quark decay with the highest perturbative accuracy.
\end{abstract}

\maketitle

\noindent \textbf{Introduction.}
The top-quark is the heaviest fermion in the standard model~(SM), and frequently
plays an important role in many extensions of the SM. Therefore, detailed
studies of its production and decay are highly desirable. Their
precise measurements at the LHC will be crucial for the understanding of
electroweak symmetry breaking and also searching for new physics. Due to its large mass,
the lifetime of the top-quark is much smaller than the typical time scale of
hadronization. For this reason, the top quark can be treated as a free particle in
good approximation, and perturbative calculations of higher order quantum corrections
to its decay rate can be performed.

Within the SM, the next-to-leading order~(NLO) QCD corrections to the top-quark decay width,
$\Gamma_t$, were calculated more than 20 years
ago~\cite{Jezabek:1988iv}. Employing the method developed
in Ref.~\cite{Czarnecki:1997fc}, the next-to-next-to-leading order~(NNLO) QCD corrections
to $\Gamma_t$ were calculated in Ref.~\cite{Czarnecki:1998qc}, in the limit
of $m_t \gg m_W$.  Later, the finite $W$ boson mass effect in the NNLO computation was taken into
account in Refs.~\cite{Chetyrkin:1999ju,Blokland:2004ye} based on the calculations of
top-quark self-energy as an expansion in $m^2_W/m^2_t$. All the previous
calculations at NNLO concentrate only on the inclusive decay width, but the differential decay
rate is also of substantial interest, especially when considering the measurement of
top-quark mass~\cite{Lancaster:2011wr} and electroweak~(EW)
couplings~\cite{Aaltonen:2012rz}. In particular, it is an important ingredient
in a fully differential calculation of top-quark pair production~\cite{Baernreuther:2012ws} and
decay at NNLO in QCD. To the best of our knowledge, such a calculation has not been finished so far
and is the subject of this Letter, in addition to the total decay width of the top quark.
%As an important cross check, the decay width of top quark at NNLO has been compared against the existing results.

\noindent \textbf{The formalism.}
We consider the SM top-quark decay,
\begin{equation}
  t \to W^+ + b + X,
\end{equation}
where $X$ represents any other parton in the final state. NNLO QCD
corrections to this process consists of three parts: two-loop virtual
contribution~($X$ contains nothing), one-loop real-virtual
contribution~($X$ contains $1$ parton), and
tree-level double real contribution~($X$ contains $2$ partons). While the amplitudes for
each part are well defined, integrals over the phase space induce
infrared singularities, which must be extracted to cancel against those from
virtual corrections in order to obtain a finite result. In particular,
the double real contribution is the primary obstacle for obtaining
fully differential NNLO corrections. {In the past decade significant
efforts have been devoted to solving this problem, and fully differential corrections
have been obtained for a number of important
processes using quite different methods~\cite{Baernreuther:2012ws,diffnnlo,Catani:2007vq}.} In this Letter, we solve this
problem for processes of heavy-to-light decay at NNLO in QCD, using a phase
space slicing method inspired by a factorization formula for heavy-to-light current in soft-collinear effective
theory~(SCET)~\cite{Bauer:2000yr}. Below
we describe our method.

To begin with, we set bottom quark mass $m_b=0$ in the NLO and NNLO QCD calculations.
Effects of finite $m_b$ are small and will be considered later as a correction to
the leading order~(LO) results. We cluster all the partons in the final state into a single jet, letting $\tau=(p_b
+ p_X)^2/m^2_t$, which measures the
invariant mass of the jet. In the limit of $\tau\to 0$, only soft radiations and~(or) radiations
collinear to the $b$ quark are allowed.  In this region, $\frac{\dd
\Gamma_t}{\dd \tau}$ obeys a factorization
formula~\cite{Korchemsky:1994jb}:
\begin{eqnarray} \frac{1}{\Gamma^{(0)}_t}\frac{\dd \Gamma_t}{\dd \tau}
&=& \mathcal{H}\left( x\equiv \frac{m^2_W}{m^2_t},\mu\right) \int\!
\dd k \, \dd m^2 J(m^2,\mu) S(k,\mu) \nn \\ &&\times\delta \left( \tau
- \frac{m^2 + 2 E_J k}{m^2_t} \right) + \cdots,
\label{eq:fac}
\end{eqnarray} where we have neglected nonsingular terms in $\tau$.
$\Gamma^{(0)}_t$ is the top-quark decay width at LO, $\mu$ is the
renormalization scale, and $E_J=(m^2_t-m^2_W)/(2m_t)$ is the energy of the
jet near threshold. $\mathcal{H}(x,\mu)$ is the hard function, which results from
integrating out hard modes of QCD in matching to SCET. It has been
calculated to NNLO in
$\als$~\cite{Bonciani:2008wf}.
$J(m^2,\mu)$ is the quark jet function with mass $m$, whose NNLO expression can be
found in Ref.~\cite{Becher:2006qw}.  It can be thought of as the
probability of finding a jet with invariant mass $m$, generated by
collinear radiations. $S(k,\mu)$ is the soft function, which describes
the probability of measuring the light-cone component of the momentum
of soft radiations $k_s\!\cdot\! n$, where $n$ is a unit light-cone
vector along the direction of the jet, to be $k$. It has also been
calculated to NNLO in Ref.~\cite{Becher:2005pd}.

Furthermore, the top-quark decay width $\Gamma_t$
can be divided into two parts:
\begin{equation}
\label{eq:gamma} \Gamma_t = \int^{\tau_0}_0\!\dd
\tau\,\frac{\dd\Gamma_t}{\dd \tau} + \int^{\tau_{max}}_{\tau_0} \!\dd\tau
\frac{\dd\Gamma_t}{\dd \tau} \equiv \Gamma_A + \Gamma_B,
\end{equation} which will be treated separately as explained below.
$\tau_0$ is a dimensionless cutoff for $\tau$, and $\tau_{max}=(1-m_W/m_t)^2$.
First, using the NNLO results for the hard function, jet function, and soft
function, we can calculate $\Gamma_A$ at NNLO, utilizing
Eq.~(\ref{eq:fac}), up to terms proportional to $\tau_0$. For
sufficiently small $\tau_0$, they can be safely neglected. The most
difficult part of the double real contributions are included in the
calculations of the jet
function and soft function.  Note that $\Gamma_A$ is infrared finite,
because the infrared divergences in the jet and soft function cancel
against those from the hard function. The spin information of the $b$ quark is
lost because spin summation has been performed in the jet function. But
polarization information of the top quark is retained, due to the fact
that soft radiations do not change spin.  In practice, instead of a
convolution form, it's more convenient to write Eq.~(\ref{eq:fac}) in a
product form:
\begin{eqnarray}
\label{eq:lapfac} && \frac{1}{\Gamma^{(0)}_t}\frac{\dd \Gamma_t}{\dd \tau} =
\mathcal{H}(x,\mu) \\ && \times\lim_{\eta \to 0}
\widetilde{j}\left( \partial_\eta + \ln\frac{m^2_t}{\mu^2},\mu\right)
\widetilde{s}\left(\partial_\eta + \ln \frac{m^2_t}{2 E_J \mu},
\mu\right) \frac{\tau^\eta}{\tau}\frac{e^{-\gamma_E
\eta}}{\Gamma(\eta)}, \nn
\end{eqnarray} where $\widetilde{j}$ and $\widetilde{s}$ are the
Laplace trasformed jet and soft function, respectively:
\begin{eqnarray} \widetilde{j}\left(\ln\frac{\nu
m^2_t}{\mu^2},\mu\right) &=& \int^\infty_0\! \dd m^2\,
\exp\left(-\frac{\nu m^2}{e^{\gamma_E}m^2_t}\right)J(m^2,\mu), \nn \\
\widetilde{s}\left(\ln\frac{\nu m^2_t}{2E_J \mu},\mu\right) &=&
\int^\infty_0 \! \dd k\, \exp\left( -\frac{2\nu
E_Jk}{e^{\gamma_E}m^2_t}\right)S(k,\mu),
\end{eqnarray} and $\tau^\eta/\tau$ should be expanded in terms of
plus distribution:
\begin{equation} \frac{\tau^\eta}{\tau} = \frac{1}{\eta} \delta(\tau)
+ \sum^\infty_{n=0} \frac{\eta^n}{n!} \left[
\frac{\ln^n\tau}{\tau}\right]_+.
\end{equation} Substituting the NNLO expansion for the hard function, jet
function and soft function into Eq.~(\ref{eq:lapfac}) gives a closed
form solution of $\dd \Gamma_t/\dd \tau$ at small $\tau$.

$\Gamma_B$ is also infrared finite. {In fact, {$\mathcal{O}(\als^2)$ contribution to it} can be obtained} from
the NLO QCD corrections to $t\to W^+ b$ plus $1$~jet, as long as
$\tau_0>0$. In our calculation, the one-loop helicity amplitudes for
this specific process are extracted from the NLO QCD corrections to single top production
associated with $W$ boson~\cite{Campbell:2005bb}.  The tree-level
helicity amplitudes are calculated with HELAS~\cite{Murayama:1992gi}.
Infrared divergences in the phase space integral of tree-level matrix elements are canceled by
adding appropriate dipole subtraction terms~\cite{Melnikov:2011qx}. {For later
convenience, we further divided the $\mathcal{O}(\als^2)$ contributions from $\Gamma_B$ into two pieces: tree-level
$t\to W^+b+2\,{\rm jets}$ plus dipole subtraction terms, $\Gamma^{(2)}_{3}$, and one-loop $t\to W^+b+1\,{\rm jet}$ plus
integrated dipole terms, $\Gamma^{(2)}_{2}$. Together with the NNLO corrections to $\Gamma_A$,
denoted by $\Gamma^{(2)}_{1}$, they add up to the full NNLO QCD corrections to $\Gamma_t$. }

{Finally, we note that throughout the calculation in this Letter, the
strong coupling constant is renormalized in the modified $\overline{\rm MS}$ scheme~\cite{Collins:1978wz},
and renormalization of masses, wave functions, and the electroweak coupling constant are 
carried out in the on-shell scheme~\cite{Denner:1990ns}.} It should be pointed out that the method used here to
calculate the NNLO corrections
is similar to the $q_T$ subtraction method of Catani and Grazzini~\cite{Catani:2007vq}.
In fact, they both employ the universality of infrared divergences and the knowledge
of resummation to facilitate the calculation.

\noindent{\textbf{Total width.}  For top-quark SM decay, the total decay width in the $G_F$
parametrization scheme~\cite{Denner:1990ns} at LO is given by
\begin{align*}
\Gamma_t^{(0)}=\frac{G_Fm_t^3}{8\sqrt{2}\pi}\Big[1-3(\frac{m_W^2}{m_t^2})^2+
2(\frac{m_W^2}{m_t^2})^3\Big],
\end{align*}
assuming CKM matrix element $|V_{tb}|=1$ and $m_b=0$. We choose
$m_W=80.385\,{\rm GeV}$, {$G_F=1.16638\times 10^{-5}\,{\rm GeV}^{-2}$}, and $m_t=173.5\,{\rm GeV}$~\cite{Beringer:1900zz},
unless specified. Other constants used in followed calculations
include $m_Z$, $\alpha_s(m_Z)$, and $m_b$, which are also chosen
as in Ref.~\cite{Beringer:1900zz}. Corrections to the LO width considered here include finite $b$ quark mass
and $W$ boson width effects, $\delta_f^b$ and $\delta_f^W$, NLO electroweak corrections,
$\delta_{EW}$, NLO and NNLO QCD corrections, $\delta^{(1)}_{QCD}$ and $\delta^{(2)}_{QCD}$,
which are defined as
\begin{align*}
\Gamma_t=\Gamma_t^{(0)}(1+\delta_f^b+\delta_f^W+\delta_{EW}+\delta^{(1)}_{QCD}+\delta^{(2)}_{QCD}),
\end{align*}
where $\Gamma_t$ is the corrected total width. In Table~\ref{ttop} we show the LO total width together
with all the corrections in percentage~(\%) for different top-quark mass values. The renormalization
scale is set to top-quark mass. Our results agree with those shown in previous literature
for finite width and mass effects~\cite{Jezabek:1988iv}, electroweak
corrections~\cite{Denner:1990ns,Eilam:1991iz}, and NLO QCD corrections \cite{Jezabek:1988iv}
%and NNLO QCD corrections~\cite{Czarnecki:1998qc,Chetyrkin:1999ju,Blokland:2004ye},
with the updated input parameters.
Especially, although using quite different method, our NNLO QCD corrections agree with the results in Ref.~\cite{Blokland:2004ye}
 within the range of  the  uncertainties of numerical calculation, which are of the order $10^{-4}$.
%Especially, although using quite different method,
%our NNLO QCD corrections agree with Ref.~\cite{Blokland:2004ye} at the level of
%about 1\% (0.02\%) of the corrections (LO width), which is remarkably good considering numerical
%uncertainties from both sides.
All the corrections are stable with respect to the top-quark mass.
\begin{table}[h!]
  \begin{center}
      \begin{tabular}{c|c|ccccc}
        \hline \hline
          $m_t$ & $\Gamma_t^{(0)}$ & $\delta_f^b$ & $\delta^W_f$ & $\delta_{EW}$ &
          $\delta^{(1)}_{QCD}$ & $\delta^{(2)}_{QCD}$ \\
        \hline
        172.5 &1.4806 & -0.26 & -1.49 &1.68 & -8.58 & -2.09\\
        \hline
        173.5 &1.5109 & -0.26 & -1.49 &1.69 & -8.58 & -2.09\\
        \hline
        174.5 &1.5415 & -0.25 & -1.48 &1.69 & -8.58 & -2.09\\
        \hline \hline
      \end{tabular}
  \end{center}
  \vspace{-3ex}
  \caption{\label{ttop}Top-quark total width at LO and corrections in percentage~(\%) from
    finite $W$ boson width, finite $b$ quark mass, and high orders, including NLO in
    EW couplings, NLO and NNLO in QCD couplings. Mass and width are shown in unit
    of ${\rm GeV}$.}
\end{table}

{As mentioned earlier, the NNLO QCD corrections can be divided into three pieces,
$\Gamma^{(2)}_i$ with $i=1,2,3$}. Each depends
strongly on the cutoff parameter $\tau_0$ up to the fourth power of $\ln\tau_0$.
While their sum should only have weak dependencies proportional to $\tau_0$, they approach
the genuine NNLO QCD corrections when $\tau_0$ is small enough. Thus in Fig.~\ref{dsdep} we
show the separate contributions to the NNLO corrections. When $\tau_0$ varies from
$10^{-3}$ to about $10^{-6}$, the separate contributions can reach as large as twice of the
LO width, while the sum remains almost unchanged at the value of about 2.1\% of the LO width.
Stability of such a large cancellation proves the validity of our NNLO calculation.
On the other hand, the NLO QCD corrections have an uncertainty of about 1.6\% of the LO width
due to the arbitrary choice of renormalization scale as shown in Fig.~\ref{scdep}, which
comes directly from running the QCD coupling constant $\alpha_s$. After adding the NNLO
QCD corrections, the scale dependence is reduced to about 0.8\%, which makes the predictions
more reliable.

\begin{figure}[h]
  \begin{center}
    \includegraphics[width=0.34\textwidth]{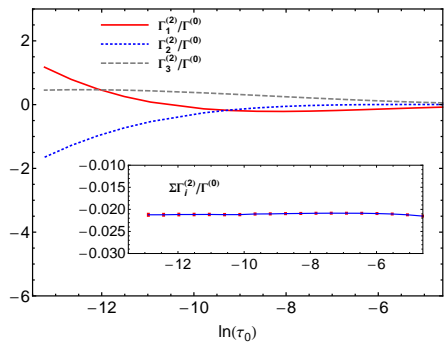}
  \end{center}
  \vspace{-3ex}
  \caption{\label{dsdep}
    Separate contributions of the NNLO QCD corrections
    and their sum as functions of the cutoff $\tau_0$, normalized
    to the LO width.}
\end{figure}

\begin{figure}[h]
  \begin{center}
    \includegraphics[width=0.38\textwidth]{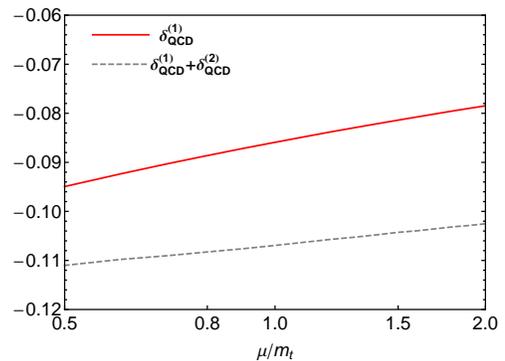}
  \end{center}
  \vspace{-3ex}
  \caption{\label{scdep}
    Renormalization scale dependence of the NLO and NLO+NNLO QCD corrections,
    normalized to the LO width at central scale $\mu=m_t$.}
\end{figure}

\noindent{\textbf{Differential distributions.} Within our framework we can calculate the fully
differential decay width of top-quark semileptonic decay $t\rightarrow W^+(l^+\nu)b$ up to
NNLO in QCD, which is not
possible for the method based on calculations of top-quark self-energy. Precise predictions
for differential distributions of top-quark decay products are of great importance, especially for
the measurement of top-quark mass~\cite{Lancaster:2011wr} and testing of the
$V-A$ structure of $tWb$ charged current~\cite{Aaltonen:2012rz}.
Below we will show several final-state distributions for $t\rightarrow W^+(l^+\nu)b$, including all
the corrections as in the total width results. {We use $e^+e^-$ $k_T$ algorithm~\cite{Catani:1991hj}
at the parton level with jet resolution threshold $y_{cut}=0.1$ for jet clustering, which is more suitable for
presentation of the results in top-quark rest frame as compared to the jet algorithms used at the LHC.}  
The shape measurements are more relevant for the experimental studies, having both small experimental
and theoretical uncertainties. Thus all the distributions shown below are normalized to unit area for comparison.
For each distribution we show results for several cases, i.e., pure LO prediction (denoted by LO1), LO predictions plus corrections
from finite $m_b$, $W$ boson width and NLO EW effects (LO2), LO2 with NLO QCD
corrections in addition, and LO2 with both NLO and NNLO QCD corrections. {In addition, We checked that the
NNLO corrections to the distributions are also stable against the cutoff $\tau_0$.}

\begin{figure}[h]
  \begin{center}
    \includegraphics[width=0.52\textwidth]{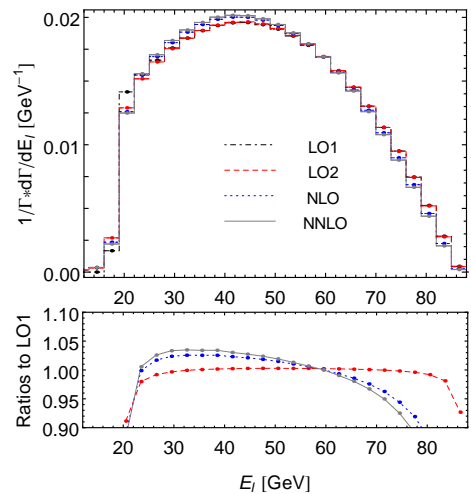}
  \end{center}
  \vspace{-5ex}
  \caption{\label{el}
   Energy distribution of the charged lepton from top-quark decay in top quark
   rest frame.}
\end{figure}

\begin{figure}[h]
  \begin{center}
    \includegraphics[width=0.52\textwidth]{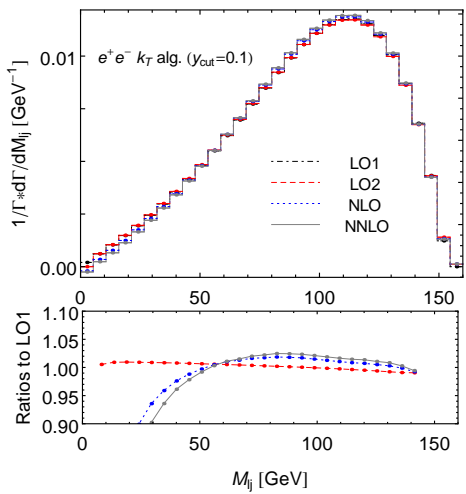}
  \end{center}
  \vspace{-5ex}
  \caption{\label{mlj}
   Invariant mass distribution of the charged lepton and hardest jet from top-quark
   decay in top quark rest frame.}
\end{figure}

In Figs.~\ref{el} to \ref{an2}, we present the charged lepton energy distribution, invariant
mass distribution of the charged lepton and the hardest jet in energy, in the top-quark rest frame,
and two angular distributions of $\cos(\theta^*)$ and $\cos(\theta_{lj})$. All of them are
normalized to unit area. $\theta^*$ are defined
in the $W$ boson rest frame as the angle between the charged lepton and the opposite of top-quark direction,
and $\theta_{lj}$ is the angle between the charged lepton and the hardest jet in top-quark rest frame. In each figure the
upper panel shows the normalized distribution while the lower panel gives their ratios
with respect to that of LO1. As we can see, the differences between LO1 and LO2 are small
in general, especially for the central region of each plot. Both the NLO and NNLO QCD corrections
push the energy and invariant mass distributions into the central region because the recoil
constituents are then massive. The NNLO corrections here are about one-fourth of the NLO
ones, similar to the results of total width. Inclusive angular distribution of $\cos(\theta^*)$
reflects the $W$ boson helicity fractions in top-quark decay, which can be also predicted
up to NNLO in QCD through top-quark self-energy calculations~\cite{Czarnecki:2010gb}.
$\cos(\theta^*)$ distribution has been extensively studied at both the Tevatron and LHC for testing
potential anomalous $tWb$ couplings induced by new physics~\cite{Aaltonen:2012rz}. By a
least $\chi^2$ fit we get the $W$ boson helicity fractions ratio as
${\mathcal F}_L:{\mathcal F}_+:{\mathcal F}_-=0.689:0.0017:0.309$ using the $\cos(\theta^*)$
distribution. The results incorporate finite $b$ quark mass and $W$ boson width effects, one-loop EW
corrections, and QCD corrections up to NNLO, and all are in very good agreement with the one shown
in Ref.~\cite{Czarnecki:2010gb}. Our calculations are more helpful for the corresponding measurements
since experimentalists can include precise corrections for the acceptance in different kinematic
regions using our results. As for $\cos(\theta_{lj})$ distribution, QCD corrections are more
pronounced there because changes of the energy spectrum also modify the distribution.

\begin{figure}[h]
  \begin{center}
    \includegraphics[width=0.52\textwidth]{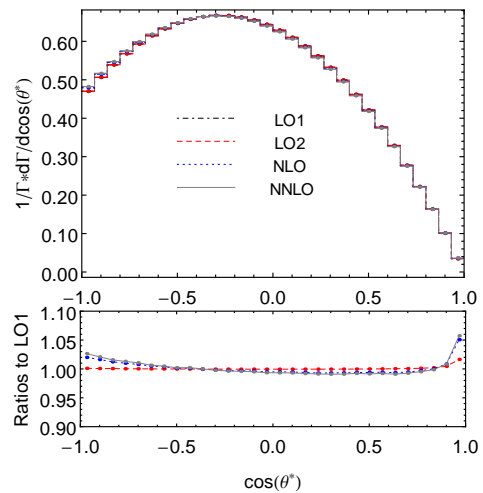}
  \end{center}
  \vspace{-5ex}
  \caption{\label{an1}
   Angular distribution of the charged lepton from top-quark decay in the $W$ boson
   rest frame.}
\end{figure}

\begin{figure}[h]
  \begin{center}
    \includegraphics[width=0.52\textwidth]{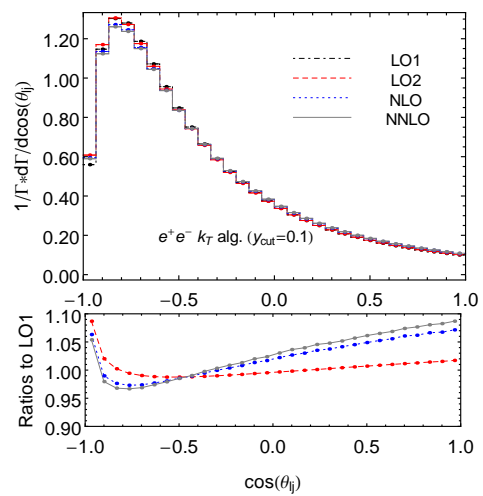}
  \end{center}
  \vspace{-5ex}
  \caption{\label{an2}
   Angular distribution of the charged lepton from top-quark decay in the top-quark
   rest frame.}
\end{figure}

\noindent{\textbf{Conclusions.} We have presented the NNLO QCD corrections to top-quark total
decay width, which do not depend on expansion in the $W$ boson mass, and fully differential
distributions of $t\rightarrow W^+(l^+\nu)b$ based on SCET. One-loop EW corrections
as well as effects from finite $b$ quark mass and $W$
boson width are also included. All together they constitute the current most precise predictions
for top-quark decay, which are helpful for top-quark mass measurement and testing of
weak charged current structure. We have implemented the calculation
into an { efficient} parton level Monte Carlo program~\cite{NNTopDec}, in which an arbitrary
infrared-safe cut can be imposed on the final state. Our calculations are complementary to the NNLO
QCD predictions for top-quark pair production~\cite{Baernreuther:2012ws}. Moreover, our method can be
widely used in studies of heavy-to-light quark decay, including $B$ meson
semileptonic decay, which will be presented elsewhere.

\begin{acknowledgments}
We appreciate helpful discussions with P. M. Nadolsky and B. Pecjak.
This work was supported by the U.S. DOE under contract DE-AC02-76SF00515, Early Career Research Award
\textrm{DE-SC0003870} by Lightner-Sams Foundation, and the National Natural
Science Foundation of China, under Grants No. 11021092,
No. 10975004 and No. 11135003.
\end{acknowledgments}

\end{document}